\documentclass[preprint,preprintnumbers,showpacs,amsmath,amssymb]{revtex4-1}
\usepackage{graphicx}
\usepackage{epsfig}
\usepackage{amssymb}
\usepackage{graphicx}
\usepackage{dcolumn}
\usepackage{bm}
\newcommand{\ba}{\begin{array}}
\newcommand{\ea}{\end{array}}
\def\br{\begin{eqnarray}}
\def\er{\end{eqnarray}}
\def\be{\begin{equation}}
\def\ee{\end{equation}}

\def\({\left(}
\def\){\right)}

\begin{document}


\title{$125$ GeV boson: A composite scalar?}

\author{A.~Doff$^1$, E.~G.~S.~Luna$^2$, A.~A.~Natale$^{3,4}$}
\affiliation{$^1$Universidade Tecnol\'ogica Federal do Paran\'a - UTFPR - DAFIS
Av Monteiro Lobato Km 04, 84016-210, Ponta Grossa, PR, Brazil \\
$^2$Instituto de F\'isica, Universidade Federal do Rio Grande do Sul, Caixa Postal 15051, 91501-970, Porto Alegre, RS, Brazil \\
$^3$Centro de Ci\^encias Naturais e Humanas, Universidade Federal do ABC, 09210-170, Santo Andr\'e - SP, Brazil\\
$^4$Instituto de F\'{\i}sica Te\'orica, UNESP 
Rua Dr. Bento T. Ferraz, 271, Bloco II, 01140-070, S\~ao Paulo - SP,
Brazil }  
 
\date{\today}

\begin{abstract}
Assuming that the $125$ GeV particle observed at the LHC is a composite scalar and responsible for the
electroweak gauge symmetry breaking, we consider the possibility that the bound
state is generated by a non-Abelian gauge theory with dynamically generated gauge boson masses and a specific chiral
symmetry breaking dynamics motivated by confinement. The scalar mass is computed with the use of the Bethe-Salpeter equation and
its normalization condition as a function of the $SU(N)$ group and the respective fermionic representation. If the fermions
that form the composite state are in the fundamental representation of the $SU(N)$ group, we can generate such light boson only 
for one specific number of fermions for each group. We address the uncertainties underlying this result, when considering the strong
dynamics in isolation.
\end{abstract}


\maketitle

\section{Introduction}

\par The ATLAS and CMS experiments at the CERN Large Hadron Collider (LHC) recently reported the discovery of a new resonance
at approximately $125$ GeV \cite{LHC}. This particle appears to be consistent with the Standard Model (SM) Higgs scalar boson, although
the data, up to now, seems to indicate an excess of events in the $\gamma\gamma$ decay branching ratio of this particle. This 
$\gamma\gamma$ decay implies that this particle is a boson, being the scalar case the simplest possibility, but we still have
a long way to determine this resonance precise quantum numbers \cite{peskin}. 

In this work we will assume that the $125$ GeV resonance is a composite scalar boson. Composite scalar bosons are known to be
formed in QCD, one example of such possibility is the elusive sigma meson \cite{polosa}, that is assumed to be the Higgs boson of QCD. 
In QCD, as shown by Delbourgo and Scadron \cite{scadron}, its mass ($m_\sigma $) is directly related to the 
dynamical quark mass ($\mu$) as
\be
m_\sigma = 2 \mu \,\, .
\label{eq1}
\ee
This relation comes out from the following relation
\be
\Sigma (p^2) \approx  \Phi_{BS}^P (p,q)|_{q \rightarrow 0} \approx \Phi_{BS}^S (p,q)|_{q^2 = 4 m_{dyn}^2 }\,\,\, ,
\label{eq2}
\ee
where the solution ($\Sigma (p^2)$) of the fermionic Schwinger-Dyson equation (SDE), that indicates the
generation of a dynamical quark mass and chiral symmetry breaking of QCD, is a solution of 
the homogeneous Bethe-Salpeter equation (BSE) for a massless pseudoscalar bound state ($\Phi_{BS}^P (p,q)|_{q \rightarrow 0}$),
indicating the existence of Goldstone bosons (pions), and is also a solution of  the 
homogeneous BSE of a scalar p-wave bound state ($\Phi_{BS}^S (p,q)|_{q^2 = 4 \mu^2 }$), which
implies the existence of the scalar (sigma) boson with the mass described above.

The BSE scalar solution depends strongly on the chiral symmetry breaking (csb) dynamics. The relation given by
Eq.(\ref{eq1}) can be modified when we consider the inhomogeneous BSE, or, in an easier approach, the homogeneous BSE 
solution associated with a normalization condition as discussed in Ref.\cite{us}, leading to lighter scalars masses. 
In particular, there are several papers in the literature discussing composite scalars, which may play the role of the
standard model Higgs boson, where it is claimed that they may have relatively low masses, as a result of a walking chiral
symmetry breaking dynamics \cite{dietrich,sannino,tuominen,us0,us1}.

The dynamics necessary to break the chiral symmetry, to form pseudoscalar and scalar bound states is connected to the behavior
of the main Green's functions of non-Abelian gauge theories (NAGT). In the QCD case the gluon propagator is a fundamental
two-point function needed to compute the SDE or BSE, and it is now known from lattice \cite{cucchieri} and SDE calculations
\cite{aguilar} that the gluon acquires a dynamical mass. This result confirms the old Cornwall's proposal that non-Abelian
gauge bosons acquire dynamical masses \cite{cornwall}, and it also imposes a severe constraint on the csb and Goldstone 
boson formation in these theories \cite{cornwall2}, even forbidden non-trivial SDE solutions leading to csb in the
case of fermions in the fundamental representation. 
A possible solution to the csb problem discussed in Ref.\cite{cornwall2} was proposed recently \cite{cornwall3}, where 
csb is intimately related to confinement, what may indeed be expected for any NAGT \cite{cornwall4}. 
We detailed the model of Ref.\cite{cornwall3} to non-Abelian gauge theories \cite{us2}
and proposed a slight modification of it in Ref.\cite{us3}. It is within this scenario that we will discuss
the possible composite origin of the boson seen at the LHC. Note that we discuss the composite scalar mass only in
the context of a pure strong interaction theory, and this mass value can be modified by radiative corrections due to
the electroweak as well as to new beyond standard model interactions necessary to generate standard fermion masses.

If this composite boson is related to the SM Higgs boson, its dynamics is also responsible for the vacuum expectation
value (vev) that promotes the electroweak gauge symmetry breaking, therefore setting the scalar boson mass to $125$ GeV, 
and using the SM vev, we may be able to infer the underlying symmetry group structure of the composite particle once
we know the csb dynamics. In the next section we discuss the csb dynamics.  This dynamics, motivated by confinement, is 
such that it may cause the decoupling of most of the degrees of freedom of the new strong interaction, and probably leaves
only such ``light" scalar boson as a reminiscent of its csb. 

The distribution of this work is the following: In Section II we discuss the chiral symmetry breaking model and how
the scalar mass comes out from the BSE and its normalization condition. In section III we explain how we can compare
the scalar mass to the data and discuss details of the group structure that appear in our mass formula. Section IV
contains a brief remark about the mass of spin $1$ composites. Section V contains our results and conclusions.

\section{CSB and the BSE}

\subsection{A model for CSB}

The standard fermionic SDE for NAGT with dynamically generated gauge boson masses in the Landau gauge is given by 
\br
&&M(p^2) = \frac{C_2 }{(2\pi)^4} \times \nonumber \\
&&\int \, d^4k \,  \frac{{\bar{g}}^2(p-k)3M(k^2)}{[(p-k)^2+m_g^2(p-k)][k^2+M^2(k^2)]} \,\, ,
\label{eq3}
\er
where $M(p^2)$ is the dynamical fermion mass ($\mu \equiv M(0)$), $C_2$ is the fermionic Casimir eigenvalue and ${\bar{g}}^2$ is
the effective charge
\be
{\bar{g}}^2(k^2)= \frac{1}{b \ln[(k^2+4m_g^2)/\Lambda^2]} \, ,
\label{eq4}
\ee
where $b=(11N-2n_f)/48\pi^2$ for the $SU(N)$ group with $n_f$ flavors, $m_g$ is the infrared dynamical gauge boson mass,
whose phenomenologically preferred value is $m_g \approx 2\Lambda $ \cite{cornwall,natale}. For fermions in the fundamental representation
of the $SU(N)$ group this coupling (${\bar{g}}(0)$) should be at least a factor $2$ larger to trigger csb \cite{cornwall2,cornwall3,haeri,
natale2,natale3}.

The approach of Ref.\cite{cornwall3} follow from a series of reasons. First, according to Ref.\cite{cornwall}, the SDE of NAGT have solutions
that minimize the energy consistent with dynamically massive gauge bosons, leading to an effective theory endowed with vortices, and these
vortices should be responsible for confinement. Lattice simulations are showing evidences for 
a relation between csb and confinement, where center vortices play a fundamental role. In the $SU(2)$ case the artificial center vortices' removal
also implies a recovery of the chiral symmetry \cite{lat1,lat2,lat3}! Objects like vortices cannot enter into the SDE at the same level of
ordinary Green's functions,
since they appear in the effective theory and must be introduced by hand. Secondly, the effective action describing confinement is an
(approximate) area-law action, which imply in an effective confining propagator, behaving as $1/k^4$, proportional to the
string tension ($K_F$) and finite at the origin due to entropic reasons, what is necessary to generate the Goldstone bosons in the csb
\cite{cornwall3}.
Therefore, we are led to introduce the following effective confining propagator in the fermionic SDE:
\be
D_{eff}^{\mu \nu}(k) \equiv \delta^{\mu \nu} D_{eff} (k); \,\,\,\,\,  D_{eff} (k)=\frac{8\pi K_F}{(k^2+m^2)^2}   \, ,
\label{eq5}
\ee
where $m$ is an entropic regulator, and the effective propagator should not at all be related do the propagation of a standard
quantum field \cite{cornwall3}.

The fermionic gap equation taking into account the dynamically massive gauge bosons and the effective confining propagator is given by
\cite{cornwall3} 
\br
&&M(p^2)=\frac{1}{(2\pi)^4}\int \, d^4k \, D_{eff} (p-k) \frac{4M(k^2)}{k^2+M^2(k^2)} \nonumber \\
&&\hspace{-0.7cm}+\frac{C_2 }{(2\pi)^4}\int  d^4k   \frac{{\bar{g}}^2(p-k)3M(k^2)}{[(p-k)^2+m_g^2(p-k)][k^2+M^2(k^2)]} \, ,
\label{eq6}
\er
where $M(p^2)=M_c(p^2)+M_{g} (p^2)$ is the dynamical fermion mass generated by the confining ($M_c(p^2)$) and one-dressed-gauge ($M_{g} (p^2)$)
boson contributions. As we remarked in Ref.\cite{us3} this equation resembles, in a different context, what we have in the successful
phenomenological quarkonium potential described by 
\be
V_F (r) = K_F r - ({4}/{3}) ({\alpha_s}/{r}), 
\label{eq6l}
\ee
were the first term is the quark confining part and the second term is the one-gluon exchange contribution. Therefore, the confining part of
Eq.(\ref{eq6}) is a reasonable phenomenological way to study csb taking into account the effective confining area law. Note that our discussion
relies heavily on QCD, although
all the facts presented here are expected to be valid for any NAGT.

The confining propagator is an effective one, and not related to a standard quantum field. Therefore it is natural to expect
a cutoff to where it can be propagated, and this point was particularly emphasized in Ref.\cite{us3}. For instance, if we think of the
phenomenological quarkonium potential that we discussed in the previous paragraph, we find a limitation up to where the linear part of the
potential is effective. We know that for $n_f =2$ quarks in the fundamental representation, lattice QCD data seems to indicate that the string
breaks at a critical distance $r_c \approx 1.25$ fm \cite{bali}. Comparatively we may set a maximum momentum $p^2 \approx K_F$ up to where the
confining part of the confining gap equation should be integrated. A discussion about separating the fermionic SDE in a confining part plus the
one-gauge boson exchange has also been  performed in a similar context in Ref.\cite{brodsky}. The solution of Eq.(\ref{eq6}) with such cutoff is
quite complicated and we will digress briefly about it in the sequence. 

Eq.(\ref{eq6}) has been solved analytically and numerically in different approximations. If we take the cutoff of both integrals of Eq.(\ref{eq6})
to infinity we can observe that $M(p^2)$ behaves asymptotically as $1/p^2$ \cite{cornwall3,us3}. But what we want is a limitation in the 
upper cutoff in the first integral. This is not easy to do, so we have set arbitrarily the upper cutoff of both integrals to a momentum scale
where the confining propagator is really effective. With this approximation the asymptotic behavior changes to a logarithmic function (details
of this calculation can be seen in Section 4 of Ref.\cite{us2}). In another approximation we assumed that the major contribution in the
momentum integration of the first integral in Eq.(\ref{eq6}) comes from the infrared region with $p,k << K_F \approx m$, expanding the
confining propagator and considering only the leading term, leads to
\br
&&M(p^2)=\frac{2}{\pi^3} \frac{K_F}{m^4} \int \, d^4k \,  \frac{M(k^2)}{k^2+M^2(k^2)} \theta (m^2-k^2) \nonumber \\
&&\hspace{-0.7cm}+\frac{C_2 }{(2\pi)^4}\int d^4k   \frac{{\bar{g}}^2(p-k)3M(k^2)}{[(p-k)^2+m_g^2(p-k)][k^2+M^2(k^2)]}.
\label{eq7}
\er
The expansion is reasonable if we compare the difference of the confining propagator (quite peaked in the infrared) with the
gauge-boson propagator (see respectively Figs.(4) and (3) of Ref.\cite{us2}). 
It is possible to verify analytically that the asymptotic behavior of this equation is logarithmic. This is easy to see because the confining
contribution has been reduced to an effective four-fermion interaction, what is equivalent to a bare mass behavior. This equation has also
been solved numerically in order to confirm the logarithmic ultraviolet behavior (the result is plotted in Fig.(9) of Ref.\cite{us2}).

In Ref.\cite{us3} it is argued that if the effective confining propagator in Eq.(\ref{eq6}) is restrained to be different from zero up to squared
momenta of order $K_F$ (or $m^2$) the effect of confinement is equivalent to the simulation of a ``bare confining" mass. This can be verified in an
extreme approach, limiting the confining propagator with Heaviside step functions and changing Eq.(\ref{eq6}) to
\br
&&M(p^2)=\frac{1}{(2\pi)^4}\int \, d^4k \, D_{eff} (p-k) \nonumber \\
&& \times \theta(K_F-k^2) \theta(K_F-p^2) \frac{4M(k^2)}{k^2+M^2(k^2)} \nonumber \\
&&\hspace{-0.7cm}+\frac{C_2 }{(2\pi)^4}\int  d^4k   \frac{{\bar{g}}^2(p-k)3M(k^2)}{[(p-k)^2+m_g^2(p-k)][k^2+M^2(k^2)]} \, .
\label{eq8}
\er
This equation can be transformed into a differential equation. Derivating the $\theta$ function we obtain a delta function and
the final effect is similar to the decoupling of the integral equations. This can be verified in the numerical evaluation of Eq.(\ref{eq8}),
which is shown in Fig.(1).

We have performed the numerical calculation of the dynamical mass for a set of constants ($K_F,m_g,\Lambda$) with values
around those typically expected for QCD.
Fig.(1) shows the dynamical mass in the case $N=3$, $K_F=0.20$ GeV${}^2$, $\Lambda=0.3$ GeV, $n_f=6$, $m_g^2=0.20$ GeV$^2$ and $C_2= 4/3$.
Note that in the sequence, always when mentioning QCD, we will work with the most usual value of the string tension $K_F=0.18$ GeV${}^2$
and a characteristic scale $\Lambda_{QCD}=\Lambda = 300$ MeV. First, the breaking is totally dominated by the confining propagator. The dynamical
mass basically
depends on the values of $K_F$ and $m$ and the infrared value is not so different from the one obtained with Eq.(\ref{eq7}). Secondly, the 
numerical result is obtained forcing the continuity of the solution, and this explains the graphics of Fig.(1): The curves are
exactly the continuous superposition of a ``constant confining mass" generated by the restrained confining propagator, plus a very
small mass, behaving asymptotically as $1/p^2$ and consistent with the value expected if we had solved the gap equation only with
the massive gauge boson propagator \cite{haeri,natale2,natale3}. In the QCD case this ``bare confining mass" can still be dressed
with the gluon exchange effects, and, stressing the discussion of Ref.\cite{us3}, we propose that the fermionic self-energy
($\Sigma(p)\equiv M(p)$) of
any NAGT are of the so called ``irregular" form and will be parameterized as \cite{us4,us5}
\be 
\Sigma (p^2) \sim \mu \left[1 + b g^2 \ln\left(p^2/ \mu^2 \right) \right]^{-\gamma }  \,\,\, ,
\label{eq12}
\ee
where\, $\mu$ \,is \,the characteristic scale of mass generation,\, 
\be
\gamma= 3c/16\pi^2 b
\label{eq12a} 
\ee
and
\be
c = \frac{1}{2}\left[C_{2}(R_{1}) +  C_{2}(R_{2}) - C_{2}(R_{3})\right]\, ;
\label{eq12b}
\ee
here $c$ is the most general Casimir operator that will appear in the case where we have a NAGT with fermions in
two different representations, $R_1$ and $R_2$, which condense and form bound states in the representation $R_3$ (in the
QCD case $c$ is reduced to the usual Casimir operator $C_2 =4/3$), $b$ is the first $\beta$ function coefficient, $g^2$ is the NAGT
coupling constant for which we assume the same expression of
Eq.(\ref{eq4}) setting $\Lambda = \mu$. 
The main feature of Eq.(\ref{eq12}), as explained in Ref.\cite{us3}, is that when this self-energy is used
in Technicolor models to compute ordinary fermion masses, the final result will depend at most
logarithmically on the gauge boson masses that connect different fermionic families. In this
case these gauge bosons can be made quite massive, and, even if they intermediate flavor changing
neutral currents, their effects will be almost decoupled from the theory, leading to viable 
phenomenological models.

We finally remark that the confining effective propagator described in Eq.(\ref{eq5}) is one possible way to model an area-law for confinement of 
fermions in the fundamental representation of a $SU(N)$ NAGT \cite{cornwall3}. This propagator, if confinement is the result of vortices, has to be
introduced by hand into the SDE, because vortices are already the result of dynamical gauge boson mass generation at a primary level.
The string breaking should also to be present in this effective theory, exactly constraining the momentum region where the confining
propagator is effective. The actual effect of confinement may still be more sophisticated than this simple model, but it does reproduce
many of the confinement characteristics learned with lattice simulations, and is a solution for csb in face of all the problems
described in Ref.\cite{cornwall2,cornwall3}. Therefore it is quite possible that confinement generates dynamical csb in NAGT,
but in a way that it looks like an explicit breaking of the chiral symmetry.

\subsection{BSE and the normalization condition}

\par The complete determination of bound states is obtained from solutions of the renormalized inhomogeneous BSE. Since the inhomogeneous BSE
is quite difficult to solve it is usual to look for the homogeneous solutions associated with a normalization condition. The BSE normalization
condition in the case of a NAGT is given by \cite{lane2}
\br
2\imath q_{\mu}= \imath^2\!\!\int d^4\!p\, Tr\left\{{\cal P}(p,p+  q)\left[\frac{\partial}{\partial q^{\mu}}F(p,q)\right]{\cal P}(p, p+  q) \right\}\nonumber \\
-\imath^2\!\!\int d^4\!pd^4\!k \,Tr\left\{{\cal P}(k,k +  q)\left[\frac{\partial}{\partial q^{\mu}}K'(p,k,q)\right]{\cal P}(p,  p+ q)\right\} 
\label{eq13}
\er
where 
$$
K'(p,k,q)  = \frac{1}{(2\pi)^4}K(p,k,q)   \,\,\, ,
$$
$$
F(p,q) =  \frac{1}{(2\pi)^4}S^{-1}(p+q) S^{-1}(p) \,\,\, ,
$$ 
where ${\cal P}(p, p + q)$ is a solution of the homogeneous BSE, $K(p,k,q)$ is the fermion-antifermion scattering kernel and $S(p)$ is
the fermion propagator. The manipulation of Eq.(\ref{eq13}) is identical to the one of Ref.\cite{us}. Skipping the algebra already
discussed in Ref.\cite{us} and identifying 
\be
G(p) \equiv \frac{\Sigma (p^2)}{\mu}
\label{eq14}
\ee
we obtain an expression for the scalar boson mass:
\br 
 M_{S}^2  =  4\mu^2&&{\Big\{}-\frac{4n_{f}N}{16\pi^2}\int d^4\!p\frac{\mu^2G^4(p){\big[}p^2 + \mu^2G^2(p){\big]^2}}{(p^2 +  \mu^2G^2(p))^4} \times \nonumber \\
 && \times \frac{1}{(p)^2}[ \gamma bg^2(p)]\left(\frac{\mu}{f'_{\pi}}\right)^2  + \nonumber  \\
&& + \,\,I^{K}(\mu, p, k, g^2) {\Big \}}. 
\label{eq15}  
\er
In Eq.(\ref{eq15}) $f'_{\pi}$ describes the composite pseudoscalar decay constant associated to $n_d$ fermion doublets and $I^{K}(\mu, p, k, g^2)$
is a higher order contribution to the BSE kernel. Working in the rainbow-ladder approximation we can neglect this contribution which is
${\cal{O}}(g^2(p^2)/4\pi)$ smaller than the first term on the right-hand side of Eq.(\ref{eq15}).

Eq.(\ref{eq12}) and Eq.(\ref{eq14}) when inserted into Eq.(\ref{eq15}), with some algebra already detailed in Ref.\cite{us}, lead to
\be 
M_{S}^2 = 4\mu^2 \left\{ \frac{bg^2(\mu)(2\gamma - 1)}{[4 + 2bg^2(\mu)(2\gamma - 1)]}\right\}.
\label{eq16}
\ee 
\par Notice that in order to have a positive mass we must have $(2\gamma - 1) > 0$, in such a way that we recover Lane's condition \cite{lane2},
i.e.
\be 
\gamma > \frac{1}{2}.
\label{eq17}
\ee

It is interesting to discuss the constraint imposed by Eq.(\ref{eq17}) on the fermion content of 
the theory. The bound state wave function, and consequently the self-energy given by Eq.(\ref{eq12}), decreases according to the value
of $\gamma$, and we must have $\gamma > 1/2$ because this is the ``hardest" expression for the wave function that we may have in field theory,
otherwise the wave function is not normalized and consistent with a localized bound state. This constraint was first obtained decades ago by Mandelstam,
was recovered in the case of gauge theories by Lane \cite{lane2}, and appears naturally in our Eq.(\ref{eq16}). If this condition is applied to QCD,
or $SU(3)$ with quarks in the fundamental representation, computing Eq.(11) and imposing $\gamma > 1/2$ we verify that the wave function is
normalized only with $n_q >5$, i.e. QCD could obey such wave function only with more than five quarks! Therefore Eq.(\ref{eq17}) will always impose
a lower limit on the number of fermions of the theories that we shall deal with.

We stress that the BSE normalization condition modify the standard result of Eq.(\ref{eq1}) only for very hard asymptotic self-energy solutions.
Otherwise it is barely possible to obtain a light composite scalar boson, because its mass is going to be twice the value of the dynamical
fermion mass,
and this one, if related to the SM vev, will lead to a quite heavy scalar boson.

\section{Group structure associated to a $125$ GeV boson mass}

Many of the $125$ GeV boson couplings observed at LHC are similar to the ones expected for the Higgs boson. Although it may even
happens that in the end this boson shall not be related to the SM symmetry breaking, the most intriguing case is the one where
it is really the responsible for the SM gauge boson masses. In this case the vev ($v$) generated by the strong interaction
is connected to the gauge boson mass through
\be
v^2= \left\langle {\bar{\Psi}}\Psi\right\rangle^{2/3} = \frac{4 M_W^2}{g_W^2} \,\, ,
\label{eq18}
\ee
where $g_W$ is the weak coupling constant, $M_W$ the charged weak boson mass, and $\left\langle {\bar{\Psi}}\Psi\right\rangle$
is the new $SU(N)$ strong NAGT condensate, whose vev is given by $v \sim 246$ GeV.

At this point we differ from the Refs.\cite{us,us1} since the dynamical mass, which appears in Eq.(\ref{eq16}), is related
to the fermion condensate (or the vev in Eq.(\ref{eq18})) through the confining propagator and consequently to the
string tension, as discussed in Ref.\cite{us2,us3}. Considering the four-fermion approximation shown in Eq.(\ref{eq7}),
and neglecting the massive one-gauge boson exchange, what is also consistent with the imposition of a momentum cutoff
of ${\cal{O}}(K_F)$ in Eq.(\ref{eq8}), the relation between the vev and the dynamical mass $\mu$ is \cite{us3}
\be
\left\langle {\bar{\Psi}}\Psi\right\rangle_R  \approx - \frac{N_R}{8\pi} \frac{m_R^4}{K_R} \mu_R \,\, .
\label{eq19}
\ee 
In Eq.(\ref{eq19}) we show the vev of fermions in the representation $R$ with dimension $N_R$, the parameter $m$ in the effective
confinement propagator is written as $m_R$, and string tension $K_R$ computed
at the scale $K_R$ \cite{us2,us3}. With Eq.(\ref{eq19}) we finally obtain the scalar boson mass
\be 
M_{S} = \frac{16\pi K_R}{N_R m_R^4} \left| \left\langle {\bar{\Psi}}\Psi\right\rangle_R \right|
 \left\{ \frac{bg^2_R(2\gamma - 1)}{[4 + 2bg^2_R(2\gamma - 1)]} \right\}^{1/2}.
\label{eq20}
\ee

The coupling $g_R^2$ in Eq.(20) is to be understood as the coupling value at the
scale where the condensate or the dynamical mass is generated, which is of the same
order of magnitude as the NAGT infrared scale. This coupling is
frozen for momenta smaller than the dynamical gauge boson mass scale, and its frozen value is
basically determined by the values of $m_g$ and $\Lambda$. Unfortunately there are no studies about how the ratio $m_g/\Lambda$ vary
for different representations. In the sequence we shall assume that this quantity does not vary strongly and the ratio is not so different
from what has been discussed in the QCD case. We will also be arguing that the dynamical mass is related to the string tension
for different representations as well as the ratio $K_R/\Lambda$ does not vary strongly with $N$ for $SU(N)$ theories.

We can now set $M_S= 125$ GeV and $\left\langle {\bar{\Psi}}\Psi\right\rangle_R \sim (246)^{3}$ GeV$^{3}$, obtaining a function involving the
variables $K_R$, $m$, $N_R$, $\gamma$, $b$ and $g^2$ for the representation $R$ computed at the scale $K_R$. There is now an important point that has been
emphasized in Refs.\cite{cornwall3,cornwall5}: Due to entropic reasons (or minimization of the energy) in order to generate the Goldstone bosons
associated to the csb we must have
\be
m^2_R \approx \mu^2_R \approx K_R \,\, .
\label{eq21}
\ee
This last equation reduces Eq.(\ref{eq20}) to an equation involving $K_R$ and quantities only dependent on the symmetry group and fermionic
content of any NAGT. This also imply that 
\be
\left\langle {\bar{\Psi}}\Psi\right\rangle_R  \approx - \frac{N_R}{8\pi}K_R^{3/2} \, ,
\label{eq222}
\ee
where the condensate is directly related to the string tension.

Considering Eq.(\ref{eq21}) we obtain the following scalar boson mass
\be 
M_{S} \approx 2 \sqrt{K_R} \left\{ \frac{bg^2_R(2\gamma - 1)}{[4 + 2bg^2_R(2\gamma - 1)]} \right\}^{1/2},
\label{eq25}
\ee 
where $K_R$ is now the typical NAGT scale that forms the composite states, $\gamma$ 
is given in Eq.(\ref{eq12a}) and obeys Eq.(\ref{eq17}), and 
\be
b= \frac{1}{(4\pi)^2}\left( \frac{11}{3} C_2(G) - \frac{4}{3}T(R)n_F (R) \right) \,\,\, , 
\label{eq26}
\ee
remembering that $C_2(R)I=T^a_RT^a_R$ and $C_2(R)d(R)=T(R)d(G)$,where $d(R)$ is the dimension of the representation $R$, while the label $G$
refers to the adjoint representation.

The string tension determining the fermion dynamical mass and the composite boson mass is now fixed by the SM condensate value. It is interesting
to recall
some properties of its value. In the representation $R$ it should be also related to the $SU(N)$ group structure and to the characteristic scale
($\Lambda$) of
the NAGT. The QCD string tension for the fundamental representation is well known from phenomenological and lattice studies, however for other
groups and representations we have to rely in lattice simulations. Lattice data for $SU(N)$ (and large $N$) seems to tell us that the ratio
$K_R/\Lambda$
is approximately constant up to order $1/N^2$ \cite{allton,lohmayer}, although this result may still be questioned \cite{greensite1} and is
connected to the way the string tensions of different representations are related, i.e. they follow a Casimir or a Sine Law scaling
\cite{greensite1,greensite2}.
Therefore, we will derive the string tension for different groups assuming that $K_F/\Lambda$ is a constant. This constant
is determined using the known value for the QCD fundamental representation string tension ($K_F=0.18$ GeV$^2$) and $\Lambda_{QCD}=300$ MeV.
We then consider Casimir scaling for the string tension
\be
K_R \approx \frac{C_R}{C_F} K_F   \,\, ,
\label{eq22}
\ee 
where $C_R/C_F$ is the ratio between the Casimir operators for the representation $R$ and the fundamental one.
For $SU(N)$ theories and a finite $N$ the Casimir scaling law must break down at some point, to be replaced by a dependence
on the $N$-ality $k$ of the representation \cite{greensite2}
\be
K_R = f(k) K_F 
\label{eq23}
\ee
This change of behavior is credited to an effect of force screening by the gauge bosons. For fermions in the adjoint representation
the $N$-ality is zero, therefore, according to Casimir scaling, the adjoint string tension is given by
\be
K_A = \frac{2N^2}{N^2-1} K_F \,\, ,
\label{eq24}
\ee
and, as a reasonable approximation, it is possible to assume $K_A \approx 2K_F$.

Finally, the composite scalar boson mass in the approximation of Eq.(\ref{eq25}) depends on the string tension, $b$ and $\gamma$ for a given
group and representation, and the value of $m_g/\Lambda$ that enters into the infrared value of the coupling constant. 
To generate our results we assume that $K_R/\Lambda$ is approximately constant for $SU(N)$ theories.
Once we have $K_F/\Lambda$ for QCD, we determine the different ratios $K_R/\Lambda$ assuming Casimir scaling,
what also give us, considering Eq.(\ref{eq21}), the relation between the dynamical mass and $\Lambda$ for a given representation.
The ratio between the gauge boson mass and $\Lambda$, based on general arguments \cite{cornwall,cornwall6}, is left to vary in the same way
it was found to vary for QCD. It is important
to remember that the ratio $m_g/\Lambda$ has a lower bound as discussed in Ref.\cite{cornwall6}, which is approximately 
given by $m_g/\Lambda \geq 1.2$, as well as we may not expect that $m_g$ is much larger than $3\Lambda$ if we assume that the NAGT
phenomenology is not too much different from what we know from QCD \cite{natale4}.

\section{A remark on a vector composite}

It is not necessary to rely on lengthy calculations to estimate the approximate composite vector meson mass in this scenario. The
vector composite mass in a NAGT with a potential like the one of Eq.(\ref{eq6l}) is heavy basically due to the spin-spin part of the
hyperfine interactions. For $S$ waves the hyperfine splitting has been determined as \cite{ei}
\be
M(^3S_1)-M(^1S_0)\approx \frac{8}{9} {\bar{g}}^2(0) \frac{|\psi (0)|^2}{\mu^2} \, ,
\label{eq27}
\ee
where $|\psi (0)|^2$ is the meson wave function at the origin, and we assumed that the fermion masses forming the meson are equal
to the dynamical mass $\mu \sim \sqrt{K_R}$. Eq.(\ref{eq27}) has been derived in the heavy quarkonium context \cite{ei}, although it seems to
work reasonably well even in the presence of light fermions (or mesons) \cite{sc}.

Assuming that no lighter composite
pseudoscalar has been seen below $125$ GeV, that ${\bar{g}}^2(0)/4\pi \approx 0.5$ \cite{cornwall3,natale4}, and that $|\psi (0)|^2 \approx \mu^3$,
what is consistent with Eq.(\ref{eq2}) we obtain the following inequality from Eq.(\ref{eq27})
\be
M(^3S_1) > (2\pi \sqrt{K_R} + 125) \mbox{GeV} .
\label{eq28}
\ee
With the dynamical fermion mass values that we obtain in this work, we can see that the vector composite is going to be a very heavy meson, whose
phenomenology will be quite model dependent.

\section{Results and conclusions} 

Our results are presented in the following Figures for a choice of $SU(N)$ groups, fermionic representations and their respective number of fermions.
In these Figures the horizontal dark gray line represents the mass (and respective uncertainty) of the boson observed at the LHC. The pale gray
vertical region is the one defining the expected values for the ratio $m_g/\Lambda$. The continuous black lines correspond to the scalar
masses computed with Eq.(\ref{eq25}) for a given number of fermions. The gauge group, number of fermions and respective representations that
we use here were chosen keeping in mind that we have to respect asymptotic freedom and the limit given by Eq.(\ref{eq17}).

In Fig.(2) we show the results for the scalar composite mass formed by fermions belonging to the $SU(N=2,3,4,5)$ gauge groups. Note that for these groups and fermions in the
fundamental representation only the theory with one specific number of fermions can generate a scalar boson of $125$ GeV. These number of fermions
are $n_f = 6,8,10$ respectively for $SU(3),SU(4),SU(5)$. In Fig.(3) we consider $n_f =6$ fermions in the fundamental representation
and verify that only in the $SU(3)$ case we can generate a composite scalar boson with a mass equal to $125$ GeV.

It is quite important to stress that the results shown here were obtained in the case of a isolated strong interaction theory. When we consider
the effect of radiative corrections, due to the electroweak and new beyond standard model interactions necessary
to generate standard fermion 
masses, the strongly generated scalar mass can be lowered by a factor of an order up to 5! 
As pointed out recently in Ref.\cite{foadi} radiative corrections induced by the effective
 scalar coupling to the top quark may decrease the scalar mass. These corrections give a contribution to the scalar mass with a negative signal typical of fermion
loops. 

In Fig.(4) we
show the case of $n_f =2$ fermions in the adjoint representation. For the groups $SU(2)$ to $SU(5)$ we do not find a solution compatible with
the LHC data. A possible solution appears only for quite large $N \,\, (>100)$.
In Fig.(5) we show the results for the two-index antisymmetric representation in the case of $SU(3)$ to
$SU(6)$ with different number of fermions, and in the $SU(5)$ (and for larger groups) no solution is found.
The scalar masses shown in the Figures result from a
delicate balance between the $\beta$ function coefficient $b$ and the Casimir
operator $c$, while we must keep the theory asymptotically free and $\gamma > 1/2$. The 
values of $g_R$ do not interfere strongly in the final result. The
scalar mass decreases with $N$ (or ``color number" $N_c$) since this leads to a larger $b$
and smaller $\gamma$ values. In the case of 2-index antisymmetric representations
the theory becomes almost conformal with a small number of fermions, the coefficient
$b$ approaches zero and the scalar mass start being more dependent on the value of
the string tension and its value increases for larger groups.
The results of Fig.(2) to (5) were obtained considering NAGT in isolation. The scalar composite masses
described in these Figures were computed under certain controllable approximations, as in the Bethe-Salpeter approach,
and we  neglected higher order corrections when discussing the effect of the BSE normalization condition in Eq.(\ref{eq15}). The results
also depend on the string tension for different representations, whereas we have a reasonable knowledge
of this quantity only for QCD. Another source of uncertainty is the value of the dynamical
gauge boson mass for different symmetry groups and with different fermionic representations. Unfortunately even for QCD we must
recognize that the dynamical gauge boson mass generation mechanism only recently started to
be studied with simulations in large lattices. Therefore, it is not necessary to stick to the face
value of $125$ GeV for the scalar mass, even in the context of an isolated strong interaction theory, uncertainties of several percent may be expected, and we
shall also comment later on the possible effect of mixing with other scalar states, what may also
introduce a large uncertainty in the calculation.

The chiral symmetry breaking solution that we discussed in Section II happens due to confinement and the dynamical mass is directly
related to the string tension. The scalar mass turns out to be strongly dependent on the string tension and its value is determined
through the SM vacuum expectation value. The other ingredient in the scalar mass formula is the value of the coupling constant in the low
energy limit, which is frozen and dependent on the ratio $m_g/\Lambda$ \cite{asn}. This dependence comes from the asymptotic behavior of the bound
state wave function. The self-energy used to compute the scalar mass (Eq.(\ref{eq12})), and the related spin $0$ wave function, is
known to occurs in the case of extreme walking theories or four-fermion dominated chiral symmetry breaking \cite{takeuchi,kondo}. The origin of
this solution in our study is totally based in confinement, appearing at an effective 
level where we may say that confinement generates a hard mass \cite{us3}. However our result for the scalar mass
is general in the sense that it does not matter what is the mechanism generating such solution, because this is the hardest asymptotic
behavior for the scalar wave function and \textit{no other behavior can lead to smaller scalar masses}.

In Section III we discussed the relation between the string tension in different representations and its relation to $\Lambda$, as well as the values of the ratio $m_g/\Lambda$. We may say that the relation between these quantities
is poorly known even in QCD, and the best evaluations for these quantities come from lattice theory
(see Ref.\cite{allton,lohmayer,greensite1,greensite2} and references therein). We tested possible variations
of the string tension for different representations with the scale $\Lambda$, and no appreciable changes compared to
the previous Figures appear in these cases.

The problem to have a light scalar associated to the SM symmetry breaking has more subtleties than the ones described here. Actually, the
 understanding of the scalar composite mass is an open problem even in the case of the QCD ``Higgs" boson, or the $\sigma$ meson (see, for instance,
a partial list of references on this subject \cite{scalar}). 

In this work we consider the possibility that the $125$ GeV boson discovered at the LHC may be a composite scalar. The homogeneous BSE tell
us that the mass of such scalar boson in one NAGT is $M_{S} \approx 2 \mu$ and of the order of the natural scale of the strong interaction
that forms the composite state. The effective scalar mass is determined from the inhomogeneous BSE, or by the homogeneous BSE plus
its normalization condition. For soft wave functions (or fermion self-energies) the normalization condition does not modify the prediction
of the homogeneous equation. However the mass is lowered when the wave function has a hard behavior. We discuss a chiral symmetry breaking
model where the wave function can decrease very slowly with the momentum. For this solution to exist the number of fermions in the theory must be
larger than a critical number given by Eq.(\ref{eq17}), otherwise the wave function is not normalized. This normalization condition is responsible for
small scalar masses.

Our results were obtained considering a pure strong interaction dynamics. The effect of radiative corrections, due to the electroweak interactions and
new beyond standard model interactions necessary to generate standard fermion 
masses, may be responsible for the decrease of the scalar composite mass, particularly due to the effect of fermion loops. This means that if the
$125$ GeV boson is indeed a composite boson it may be necessary a precise engineering of different interactions to explain its mass.

There are important points that remain to be answered in this problem, as, for instance, the effect of the next order corrections to the BSE
normalization condition. However there are also questions that may be answered soon by lattice simulations: Small composite scalar masses can
be obtained as a consequence of a wave function that decreases slowly with the momentum, they imply in a constraint on the number of fermions
depending on the group and fermionic representation. If this constraint is not obeyed probably the scalar masses tend to be large, because
the chiral symmetry breaking mechanism is different from the one discussed here with softer wave functions and fermionic self-energies. Therefore, 
it will be quite interesting to have simulations of NAGT providing the scalar mass values for different groups and fermionic representations, which
may even be a test of the chiral symmetry breaking dynamics.

\section*{Acknowledgments}

We thank A. C. Aguilar for discussions and help with the numerical calculation and M. A. Kneipp for discussions. This research was partially
supported by the Conselho Nacional de Desenvolvimento Cient\'{\i}fico e Tecnol\'ogico (CNPq).

\newpage

\begin{figure}[!h]
\centering
\hspace*{-0.2cm}\includegraphics[scale=0.70]{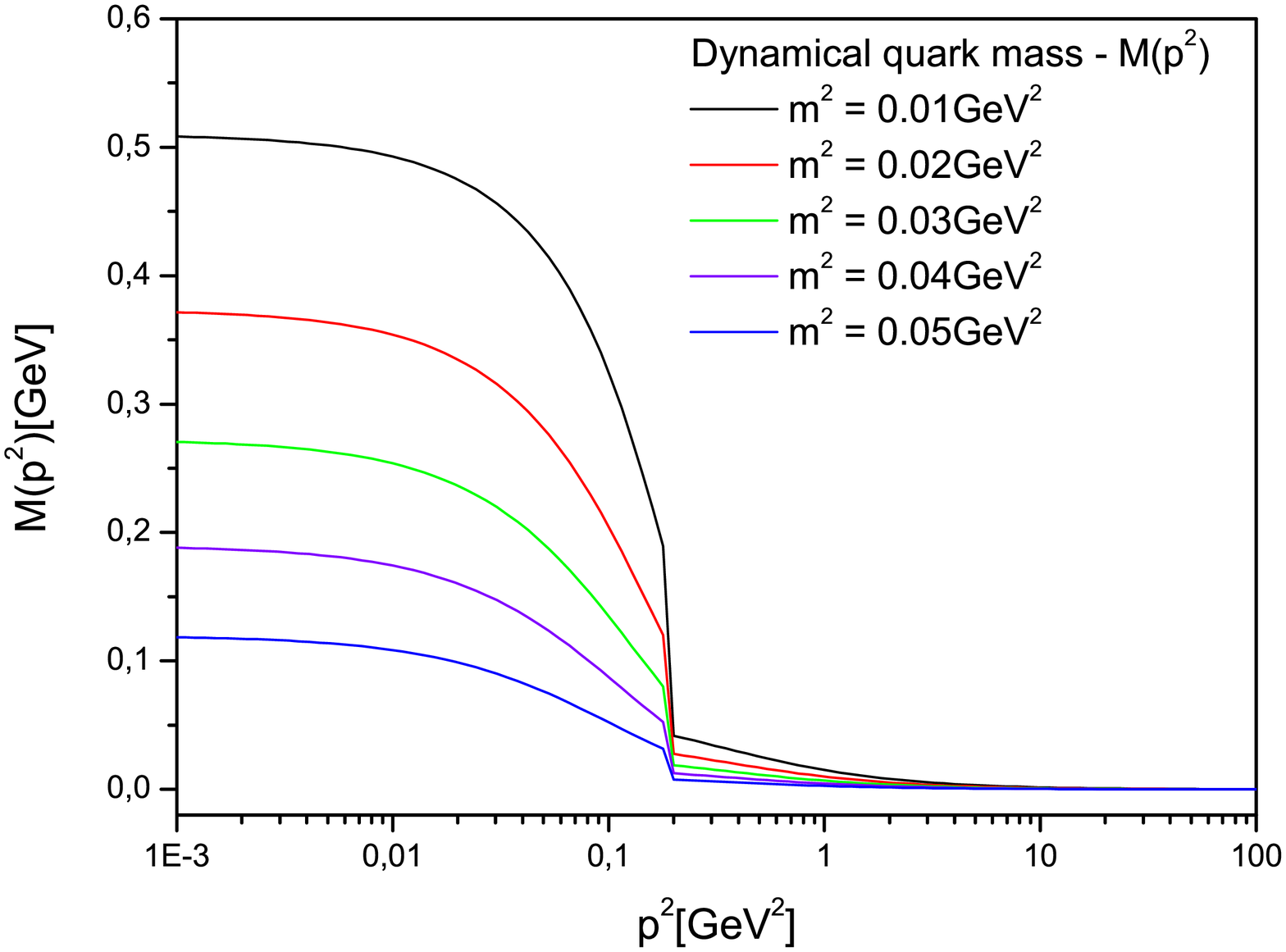}
\label{fig1}
\caption{Dynamical quark mass obtained with the numerical calculation of the equation (\ref{eq8}).}
\end{figure}

\begin{figure}[!h]
\centering
\hspace*{-0.2cm}\includegraphics[scale=0.70]{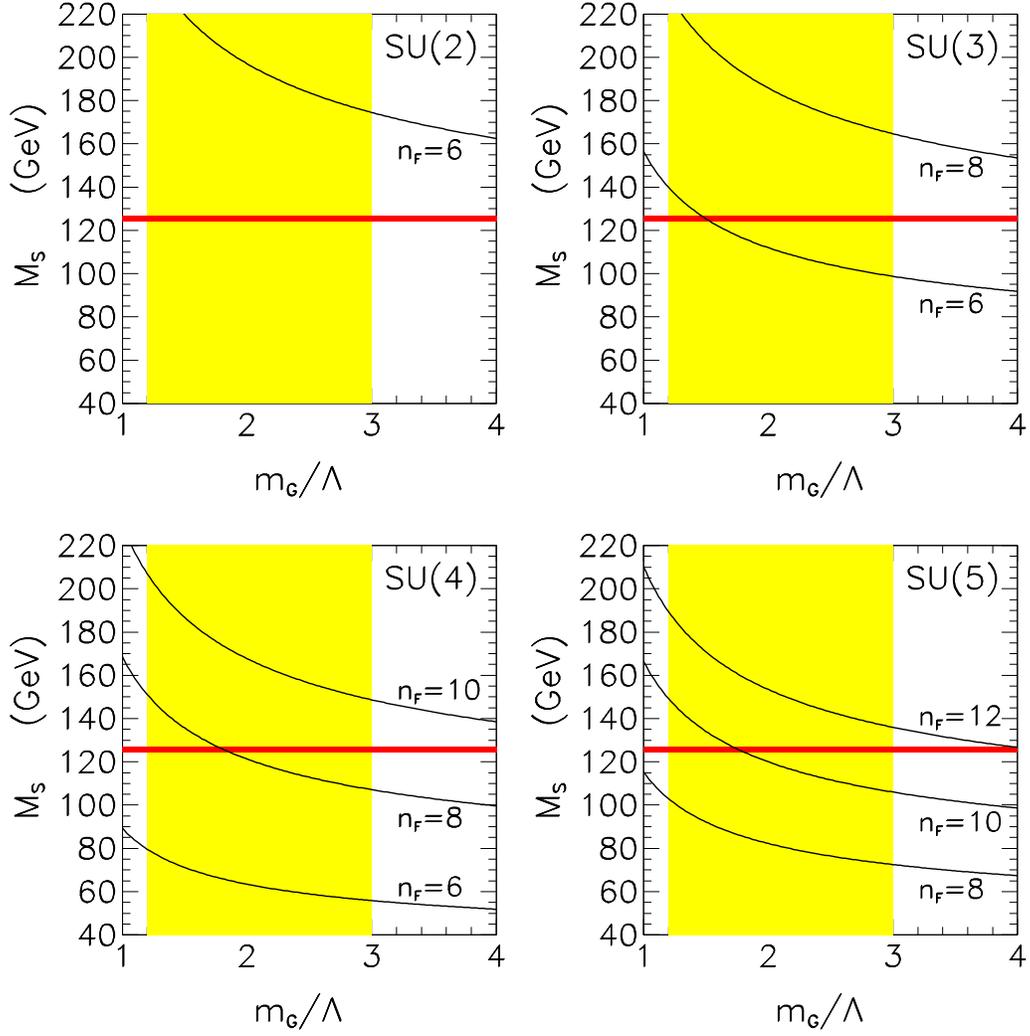}
\label{fig2}
\caption{Scalar boson mass $M_{S}$ calculated using the $SU(N=2,3,4,5)$ gauge group in the fundamental representation with different numbers of
Dirac fermions.}
\end{figure}

\begin{figure}[!h]
\centering
\hspace*{-0.2cm}\includegraphics[scale=0.70]{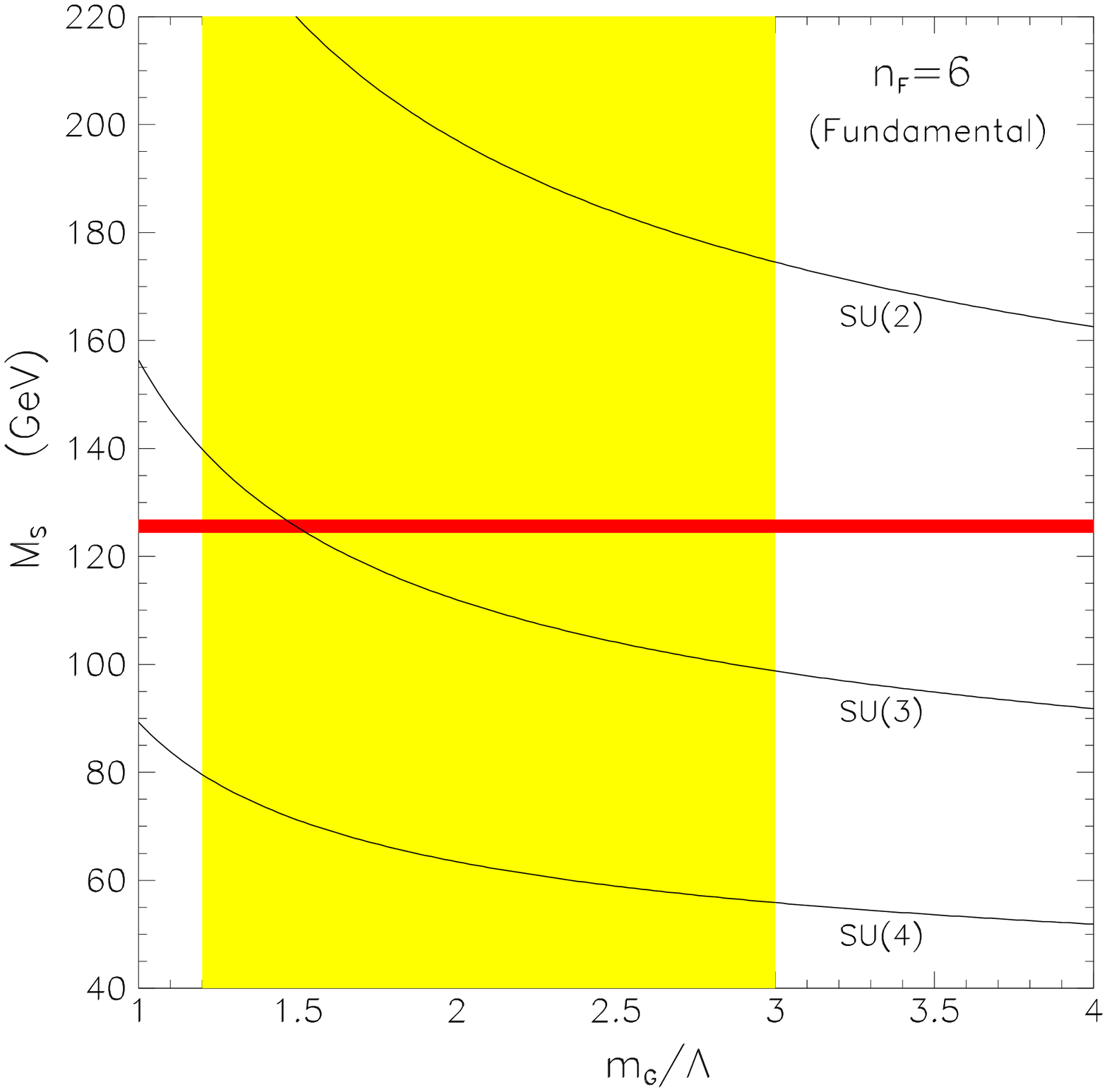}
\label{fig3}
\caption{Scalar boson mass $M_{S}$ calculated using the $SU(N=2,3,4)$ gauge group in the fundamental representation with the number
of Dirac fermions $n_{F}$ set at $6$.}
\end{figure}

\begin{figure}[!h]
\centering
\hspace*{-0.2cm}\includegraphics[scale=0.70]{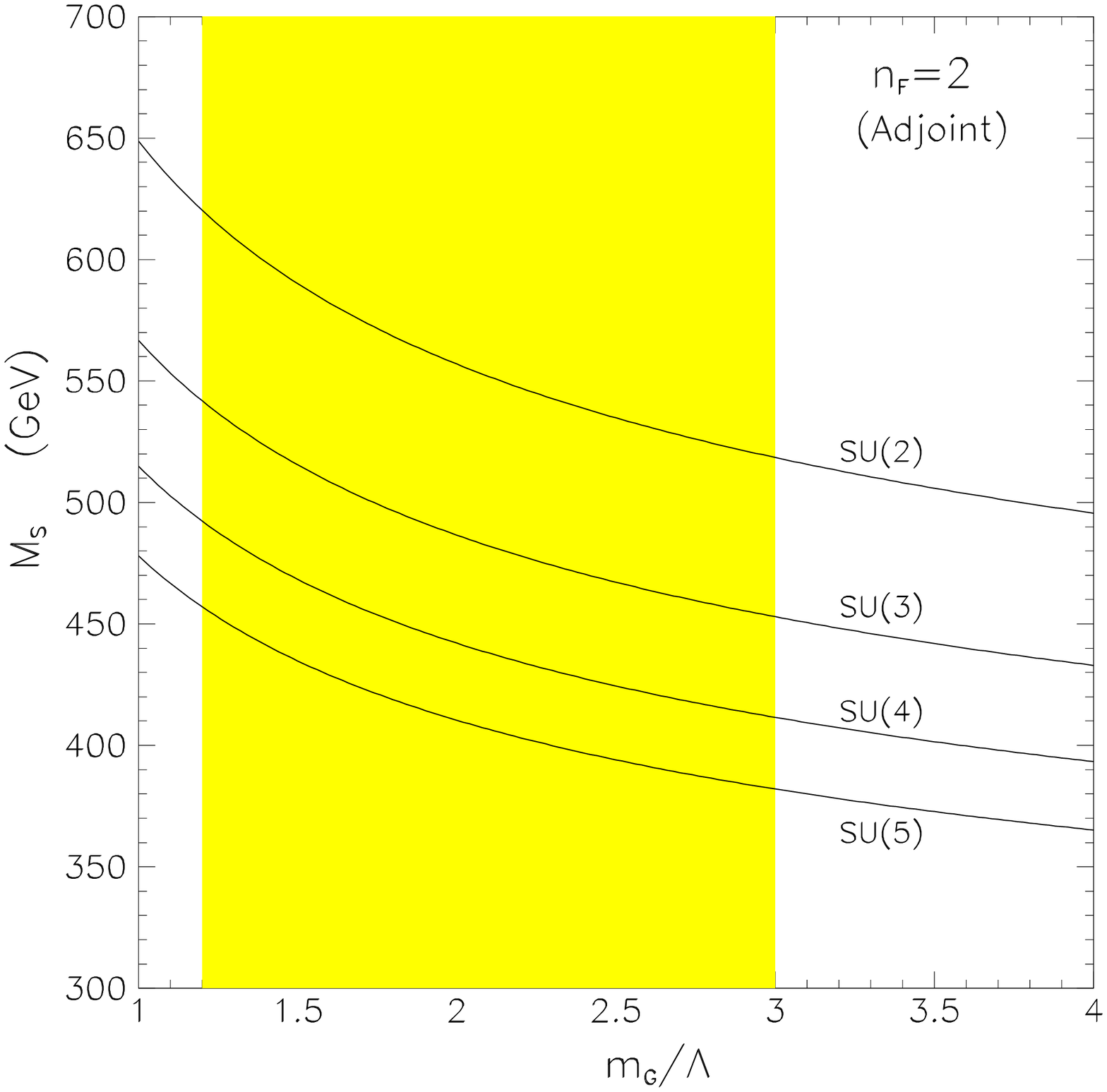}
\label{fig4}
\caption{Scalar boson mass $M_{S}$ calculated using the $SU(N=2,3,4)$ gauge group in the adjoint representation with the number
of Dirac fermions $n_{F}$ set at $2$.}
\end{figure}

\begin{figure}[!h]
\centering
\hspace*{-0.2cm}\includegraphics[scale=0.70]{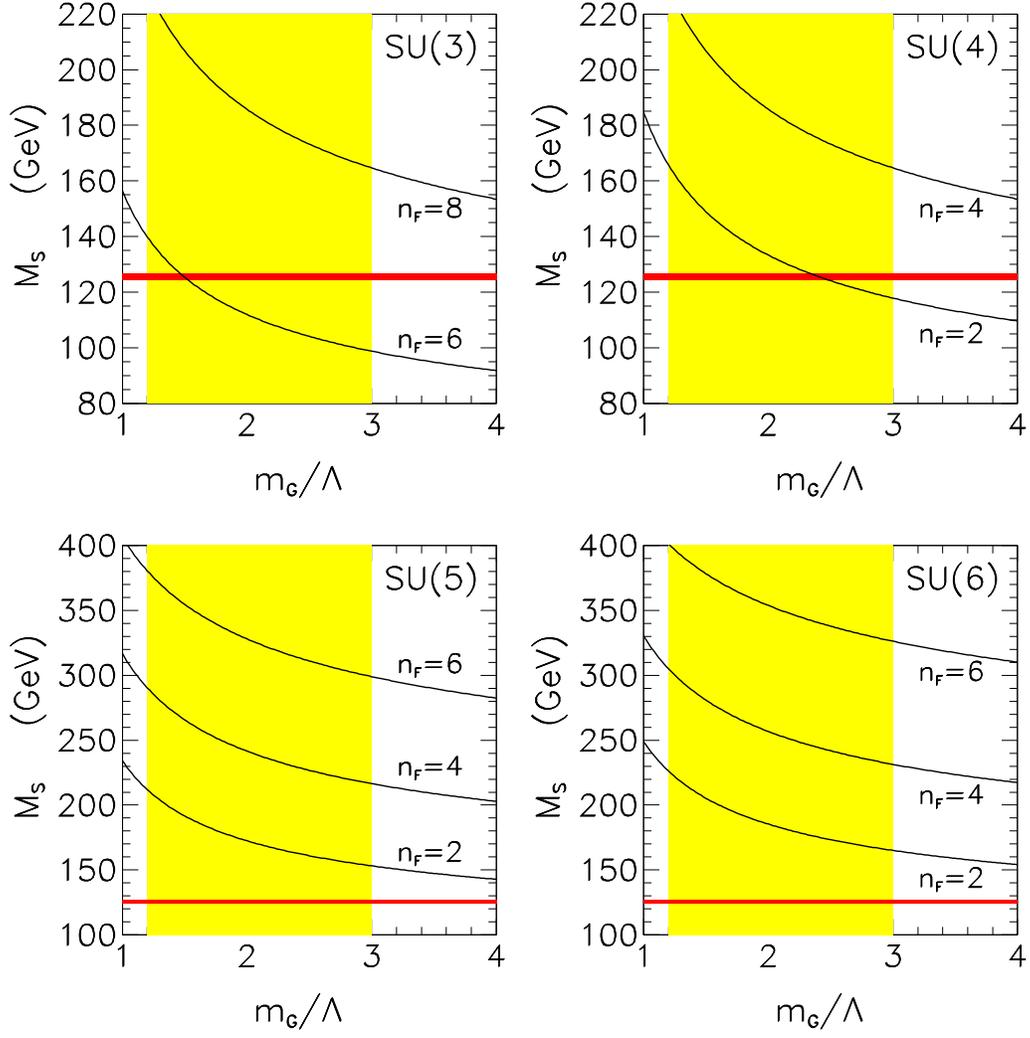}
\label{fig5}
\caption{Scalar boson mass $M_{S}$ calculated using the $SU(N=3,4,5,6)$ gauge group in the 2-index antisymmetric representation with different
numbers of Dirac fermions.}
\end{figure}

\end{document}